\documentclass[fleqn,10pt]{iopart}
\usepackage[utf8]{inputenc}
\usepackage[T1]{fontenc}
\expandafter\let\csname equation*\endcsname\relax

\expandafter\let\csname endequation*\endcsname\relax

\usepackage{
	amsmath,
	amsbsy,amsfonts,amstext,txfonts,
	amssymb,
	rotating,dcolumn,graphics,array,multirow,sidecap,mathrsfs,textcomp,%gensymb,
	graphicx,units,hyperref}
\usepackage{color,booktabs}
\usepackage{lscape}
\usepackage{adjustbox}
\definecolor{violet}{rgb}{0.56, 0.0, 1.0}
\DeclareUnicodeCharacter{2273}{\ensuremath{\gtrsim}}

\newcommand{\sona}{}

\usepackage{color}

\usepackage{geometry}  % To customize the page layout
\usepackage{fancyhdr}  % For customizing headers
\usepackage[authoryear]{natbib}
\renewcommand{\harvardurl}{\textbf{URL:} \url}
\begin{document}

\vspace{-2.0cm}
\title{Generalized, sublethal damage-based mathematical approach for improved modeling of clonogenic survival curve flattening upon hyperthermia, radiotherapy, and beyond}

% Custom header using fancyhdr
\pagestyle{fancy}
\fancyhead{}  % Clear current header
%\fancyhead[R]{\textit{Generalized, sublethal damage-based mathematical approach for improved modeling of clonogenic}\\ \textit{ survival curve flattening upon hyperthermia, radiotherapy, and beyond} \\ }  
\renewcommand{\headrulewidth}{0pt}
\fancyhead{}  % Clear current header to remove title and any other header content

\author{Adriana M. De Mendoza$^{1,2,*,+}$, So\v{n}a Michl\'ikov\'a$^{2,3,+}$, Paula S. Castro$^5$, 
	Anni G. Mu\~noz$^1$, Lisa Eckhardt$^{2,6,7}$, Steffen Lange$^{2,4,++}$, 
	and Leoni A. Kunz-Schughart$^{2,8,*,++}$}

\address{$^1$ Pontificia Universidad Javeriana, Physics Department, Bogotá, 110231, Colombia}
\address{$^2$ OncoRay—National Center for Radiation Research in Oncology, Faculty of Medicine and University Hospital Carl Gustav Carus, TUD Dresden University of Technology, Helmholtz-Zentrum Dresden—Rossendorf, 01307 Dresden, Germany}
\address{$^3$ Helmholtz-Zentrum Dresden-Rossendorf, Institute of Radiooncology - OncoRay, 01328 Dresden, Germany}
\address{$^4$ DataMedAssist Group, HTW Dresden—University of Applied Sciences, 01069 Dresden, Germany}
\address{$^5$ Universidad Distrital - Francisco José de Caldas, Bogotá, 111611, Colombia}
\address{$^6$ Core Unit for Molecular Tumor Diagnostics (CMTD), National Center for Tumor Diseases Dresden (NCT/UCC), Germany: German Cancer Research Center (DKFZ), 69120 Heidelberg, Germany; TUD Dresden University of Technology, Helmholtz-Zentrum Dresden—Rossendorf (HZDR), 01307 Dresden, Germany}
\address{$^7$ German Cancer Consortium (DKTK), Partner site Dresden, and German Cancer Research Center (DKFZ), 69192 Heidelberg, Germany}
\address{$^8$ National Center for Tumor Diseases Dresden (NCT/UCC), Germany: German Cancer Research Center (DKFZ), 69120 Heidelberg, Germany; TUD Dresden University of Technology, Helmholtz-Zentrum Dresden—Rossendorf (HZDR), 01307 Dresden, Germany}

\eads{\mailto{a.demendoza@javeriana.edu.co}, \mailto{leoni.kunz-schughart@oncoray.de}}

\footnote{+ Shared first authorship.}
\footnote{++ Shared last authorship.}

\begin{abstract}{}\\
	
\noindent \textbf{Objective:} Mathematical modeling can offer valuable insights into the behavior of biological systems upon treatment. Different mathematical models (empirical, semi-empirical, and mechanistic) have been designed to predict the efficacy of either hyperthermia (HT), radiotherapy (RT), or their combination. However, mathematical approaches capable of modeling cell survival from shared general principles for both mono-treatments alone and their co-application are rare. Moreover, some cell cultures show dose-dependent saturation in response to HT or RT, manifesting in survival curve flattenings. An advanced survival model must, therefore, appropriately reflect such behavior.

\noindent \textbf{Approach:} We propose a mathematical approach to model the effect of both treatments based on the general principle of sublethal damage (SLD) accumulation for the induction of cell death and irreversible proliferation arrest. Our approach extends Jung’s model on heat-induced cellular inactivation by incorporating dose-dependent recovery rates that delineate changes in SLD restoration.

\noindent \textbf{Main results:} The resulting unified model (Umodel) accurately describes HT and RT survival outcomes, applies to simultaneous thermoradiotherapy modeling, and is particularly suited to reproduce survival curve flattening phenomena. We demonstrate the Umodel’s robust performance (R2 ≳0.95) based on numerous clonogenic cell survival data sets from the literature and our experimental studies.

\noindent \textbf{Significance:} The proposed Umodel allows using a single unified mathematical function based on generalized principles of accumulation of sublethal damage with implemented radiosensitization, regardless of the type of energy deposited and the mechanism of action. It can reproduce various patterns of clonogenic survival curves, including any flattening, thus encompassing the variability of cell reactions to therapy, thereby potentially better reflecting overall tumor responses. Our approach opens a range of options for further model developments and strategic therapy outcome predictions of sequential treatments applied in different orders and varying recovery intervals between them.

\end{abstract}

%=======================================================================================================================================================================================
	\flushbottom
%	\maketitle
	\setcounter{secnumdepth}{5}

\section*{Introduction}
Radiotherapy (RT) continues to be the second most common medical intervention prescribed to more than half of the patients diagnosed with cancer~\cite{Chandra,Begg}. Even though RT is very potent in reducing the tumor mass, the dose required to cure the patient can often not be applied due to high radiotoxicity in the adjacent normal tissues. Therefore, combining RT with local or selective radiosensitizing moieties can be crucial for better (curative) therapeutic outcomes. Heat is one of the most potent radiosensitizers, achieving thermal enhancement ratios (TER) as high as 8.0 in \textit{in vivo} studies~\citep{Overgaard2013,Overgaard,Mei,Elming}. A growing number of recent topic-related publications and, to this date, 55 reported active or completed clinical trials evidence the revival of thermoradiotherapy (TRT) in cancer research and treatment~\cite{CT}, mainly motivated by state-of-the-art advances in a localized, spatially, and temporally controlled heat application~\cite{Kok2,Georgios,Schupper,lian,Cheng,Paulides,Kang}.\\ 

Tumor response to treatment and prediction of patient survival are both based on cell survival models, which are translated by mathematical approaches or simulations into tumor control probabilities (TCPs)~\cite{Hillen, Naqa, Borkenstein}. Different mathematical approaches (empirical, semi-empirical, and mechanistic) have been designed to model the efficacy of RT, HT, and their combination. Only mechanistic models allow the development of new therapeutic approaches based on a hypothesis-driven understanding of synergistic biological effects. Furthermore, as soon as biological treatment planning is integrated into the clinical routine of TRT, suitable and valid mechanism-based mathematical models are required for a personalized treatment design. For ionizing radiation, the LQ model offers one of the most robust approaches to predicting the survival fraction in RT treatments~\cite{Brenner}. Although initially developed as an empirical approach, it has since been given mechanistic interpretations related to the probabilities of radiation-induced DNA damage~\cite{Chadwick}. However, the validity of these interpretations is still a matter of debate. Most importantly, the LQ model is not suited for mechanistically describing the effects of hyperthermia, as DNA breaks are usually not induced by conventional HT regimes (40-50 $^\circ$C)~\cite{Lepock}. Likewise, other mechanistical approaches~\cite{Tobias,Friedrich4,Friedrich2}, are also explicitly suited for RT. Notably, many of the models referenced above encounter limitations at low-survival or high-dose regimes, where the cell kill is overestimated, e.g., at the tail of the logarithmic survival curves, which are typically straight for mono-RT~\cite{McMahon}. Some alternative mathematical models deal with this issue~\cite{Bruningk1,Park,Scheidegger,Bodgi}. Similarly, the various mathematical strategies to model HT-induced cell killing are not readily applicable to RT responses, and there is poor consensus on the underlying biology~\cite{Pearce,Jung,Roti,Uchida,Mackey}.\\ 
\vspace*{-0.3cm}

Regarding TRT, a component reflecting the HT-induced radiosensitization has to be implemented in addition to the cumulative cell killing by the two individual therapies. Several proposed models provide good empirical approximations ($R^2 \gtrsim 0.95$), as summarized in our previous publication~\cite{DeMendoza}. A comprehensive overview of the major mathematical approaches currently available for modeling cell survival upon RT, HT, and TRT is given in the supplemental information (section SI.1) Tables~SI.1a-c. Evidentially, mathematical approaches capable of modeling cell survival from shared general principles for both mono-treatments alone and for their co-application are rare, in particular, mechanistic models that can contribute to a better understanding of the synergistic effects of the combined therapy. The available models poorly reproduce those clonogenic survival curves that are substantially affected by thermal or radiation-dose dependent changes in cell population recovery, e.g., among others, due to alterations in the kinetics of DNA repair, protein refolding, metabolic adaptations, or the existence of resistant subpopulations. Such effects are expressed as further straightening or flattening of the clonogenic cell survival curves and occur in various cell types subjected to ionizing photon or heavy ion radiation and upon moderate HT. The phenomenon of survival curve flattening has, for example, been observed in the radiation response of lymphocytes~\cite{Pag2023,PhaCouThaVal2024}, a gold standard \textit{in-vitro} assay for determining individual radiosensitivities in humans. Therefore, an advanced survival model must appropriately reflect such behavior.\\
\vspace*{-0.3cm}

We advocate that a meaningful mechanistic model must allow the accumulation of sublethal damage (SLD) to encompass the effects of heat and irradiation in mono- and combination treatments. Only a few models incorporate this feature~\cite{Jung,Uchida,Scheidegger}. However, these models are suited only for combined TRT and cannot emulate the respective mono-therapies. The same is true for our recently introduced model for simultaneous TRT, which is based on thermodynamic principles, where radiosensitization is defined as an accumulation of HT-induced SLD~\cite{DeMendoza}. In the present study, we propose Jung's model as an integrated framework to (i) describe the effect of RT, HT, and TRT from shared underlying principles and (ii) reproduce survival curves presenting saturation of the cell-killing effect with increasing dose. Jung's model stipulates that cells lose their reproductive capacity due to damage accumulation in discrete stages without reliance on any specific mechanistic principle. However, it does not contain components delineating changes in SLD restoration rates. Hence, it can neither reflect the effects of HT on the repair of RT-induced DNA damage, a fundamental phenomenon in HT-induced radiosensitization, nor the flattening of the survival curve.\\
\vspace*{-0.3cm}

In the newly presented approach, we modified Jung's model~\cite{Jung} by incorporating a dose-dependent rate of SLD recovery. The recovery rate is modeled as an effective enzymatic reaction, accounting for all possible restoration mechanisms of accumulated non-lethal damage at the cellular or the population level. The radiosensitizing effect of HT is mathematically implemented by reducing the repair rates for the RT-induced damage upon HT. These modifications improve the accuracy in modeling dose-response relationships. We conducted a thorough and comprehensive testing of our “unified” model (Umodel) on various cell survival data from the literature and our experimental data in a panel of human head and neck squamous cell carcinoma (HNSCC) cell lines, where HT and RT were applied individually and in combination. In the mono-treatment cases, we compare the goodness-of-fit of the Umodel to the standard LQ, the LQC, and Jung's models, yielding comparable or, in selected cases, preferable results. The latter is particularly noticed when therapeutic phenomena such as thermal adaptation or high-dose radioresistance are observed. Here, we demonstrate the ability of the Umodel to reproduce straightening and flattening survival curves with superior fidelity. Based on our findings, it may be beneficial to consider using the Umodel for predicting cell survival upon TRT in biological treatment planning. The report concludes by discussing the advantages and limitations of the newly developed Umodel, providing an outlook, and suggesting possible future uses and improvements.\\

\section*{Methods}

\subsection*{Biological Terminology }

The key biological terms used in this work are specified as follows (adapted from De Mendoza et al. 2021~\cite{DeMendoza}):

\begin{itemize}
\item {\bf Sublethal cell damage:} Any non-lethal deterioration of cellular processes, regardless of origin and kinetics, that advances the cell toward a dead state. In the Umodel, the sublethal damage accumulates with rate $r$.
\item {\bf Sublethal damage repair:} Any cellular process, regardless of underlying biological mechanism and kinetics, leading to restoration of the sublethal damage. In the Umodel, repair is defined only as a rate q(t) with which the cell 'returns' to the previous compartment.
\item {\bf Cell kill (``dead state/compartment''):} From the radiotherapeutic perspective, cells are considered to be dead (killed) when they have lost their reproductive capacity, i.e., they are no longer able to divide and become replication-incompetent. It encompasses cells losing their membrane integrity and cells truly dying by apoptosis, necrosis, or other mechanisms, but also living cells undergoing terminal differentiation, permanent cell cycle arrest, or senescence. This type of cell kill leads to the control of a malignant disease, independent of the underlying process. In the Umodel, the dead state is a final compartment reached when a cell cannot accumulate more sublethal damage.
\item {\bf Cell survival (``alive state/compartment''):} A cell is considered to survive if it remains replication-competent, i.e., when retaining reproductive capacity after treatment. In the Umodel, the cell is alive in all compartments $(n)$ from $n=0$ to $n_\text{max}$.
\end{itemize}

\vspace{-0.2cm}

\subsection*{Development of the ``Umodel'' }
\vspace{0.2cm}

\subsubsection*{Original Jung's model}

\begin{figure}%[H]
        \centering
        \includegraphics[scale=0.7]{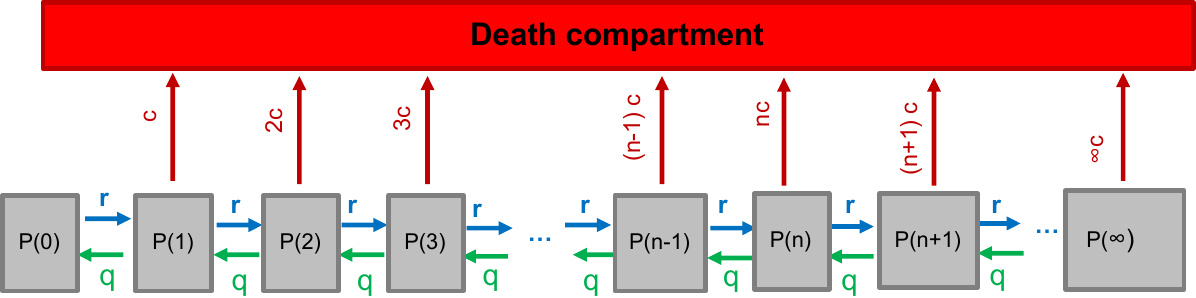}
        \caption{\textbf{Compartment model of cell killing by HT.} The original Jung's model~\cite{Jung} is represented by the scheme without regression rates $q$ (green arrows). Our proposed modification called Umodel includes regression rates $q$ (green arrows). In the Umodel (as in Jung's), compartments constitute a basis of possible states of the cell population, and $P(n)$ is the  proportion (probability) of the population in the $n$-th state.}
        \label{Jung}
\end{figure}

Jung's model considers an infinite number of SLD accumulation stages, also called compartments. At the $n$-th stage, a fraction of surviving cells endures $n$ non-lethal lesions. The probability that the cell is in the $n$-th compartment, also reflecting the fraction of cells in the compartment, is given by the solution of the detailed balance equation

\begin{equation}\label{BE}
        \frac{dP_n(t)}{dt}=-rP_n(t)-ncP_n(t)+rP_{n-1}(t),
\end{equation}

which is a time-continuous Markov chain. It describes the time evolution of the probability at the $n$-th compartment in a way that cells can advance in a sequence of SLD with a rate r, or escape to death with a rate $nc$, proportional to the stage of non-lethal damage accumulation (Figure~\ref{Jung}). Here, $r$ is defined as the rate of SLD accumulation, and $c$ is the rate of damage fixation, which refers to processes that prevent further damage repair in a non-reversible manner. Thus, the state of the cell population is given by the probability vector $\vec{P}(t)=(P_0(t),P_1(t),P_2(t),\ldots,P_n(t),\ldots)^T$, whose $n$-th element is the probability that the cell is in the $n$-th compartment at time $t$. Jung's original approach, initially proposed to model the effect of heat on \emph{Chinese hamster ovary} (CHO) cells \textit{in vitro}, very well reproduces the HT outcome~\cite{Jung}. In this case, the advance rate $r$ between consecutive compartments in the chain of non-lethal damage is constant.\\

Equation (\ref{BE}) is solved under the following boundary conditions:

\begin{enumerate}
\item In the first compartment ($n = 0$), cells are in their initial undamaged state, a pivotal starting point. They are only damaged and move forward after the onset of the treatment $\frac{dP_0(t)}{dt}=-r_0P_0(t)$. %, it is $P_0(t)=e^{-(a+r_0)t}$.
\item At $t=0$, immediately before the start of treatment, $P_{n=0}(0)=1$ and $P_n(0)=0$ (for $n\geq 1$). Accordingly, the initial condition can be written as the state vector $\vec{P}(t=0)=(1,0,0,0,\ldots,0)^T$.
\item The concept of cell killing is straightforward: it requires damage. Hence, the cell undergoes at least one stage of damage before dying.
\end{enumerate}

Furthermore, the evolution of the state vector is expressed as $\frac{d\vec{P}(t)}{dt}=\hat{A}\vec{P}(t)$, where the elements of the transition matrix $\hat{A}_{ij}$ define the influx rate from $n = j$ to $n = i$, and the diagonal elements are the net flux at each stage. The survival probability is given by the probability of being in any of the non-lethal damage compartments:
\vspace{-0.2cm}
\begin{equation}\label{Gsurv0}
        \mathcal{S}(t)=\sum_{n=0}^{\infty} P_n(t).
\end{equation}

In Jung's original model, the applied thermal dose is proportional to the treatment time for a fixed heat intensity (determined by the temperature) $D=\dot{D}(T)t$. When cellular damage is inflicted, biological responses are triggered and may take seconds to days to complete~\cite{Frankenberg,Alberts}. In HT treatments with conventional heat sources, the dose rate is the pace of heat deposition and is related to temperature. The treatment temperature is usually set in HT experiments, i.e., the exposure time determines the applied total thermal dose. In the case of RT in preclinical, experimental settings, the dose rate is usually pre-determined by the power of the irradiation device; the desired total dose is then administered by adjusting the exposure time. Thus, time refers to treatment duration, while damage advancement and fixation rates depend on the dose rate $\dot{D}$. For the dose rates typically used in external-beam radiotherapy, the exposure times (duration of single RT treatment or fraction) are relatively short, and the concomitant rapid induction of DNA damage is generally counteracted by repair processes taking place on a different time scale~\cite{RBbook}. Hence, cell recovery or (reproductive) death do not occur during the short treatment interval of RT; the survival outcome thus depends, in most cases, on the applied irradiation dose but not the dose rate.

\subsubsection*{Regression rate in Jung's model}

Jung's model does not include the possibility of regressing in the chain of SLD. We incorporated this feature to describe possible tissue adaptation or recovery. We considered two possibilities for the regression rate $q$, dependent either on the stage of damage $n$ or the treatment time $t$, to ensure we cover all potential scenarios. When an $n$-dependent regression rate $q_n$ is included in Jung's model, the net advance rate $r_n=r-q_n$ changes with the level of SLD. The general solution of $P_n(t)$ is presented in Section~SI.2 of the supplementary information. We tried different stage-dependent functions for $q_n$ (cf. Eqs. ((SI.4)-(SI.6)), but the results yielded no improvement over the original Jung's model.\\

After ruling out a stage-dependent regression, we examine a dose-dependent regression function, i.e., treatment time $t$ for a fixed $\dot{D}$. Including a time-dependent function $q(t)$ also makes the transition matrix $\hat{A}\rightarrow \hat{A}_t$ time-dependent. In this particular case, the detailed balance equation reads
\vspace{-0.1cm}
\begin{equation}\label{bal}
        \frac{dP_n(t)}{dt}=-r(t)P_n(t)-ncP_n(t)+r(t)P_{n-1}(t) \text{ ,}
\end{equation}

The net advance rate in the SLD chain in Eq.~(\ref{bal}) is then given by

\begin{equation}\label{rot}
  r(t)=r-q(t).
\end{equation}

with $r$ as the original rate of SLD accumulation. It is worth
mentioning that Eq.~(\ref{bal}) can be written in the compact form
$\frac{d\vec{P}(t)}{dt}=\hat{A}(t)\vec{P}$. The transition matrix is
an infinite-dimensional square matrix, containing only diagonal
elements, given by ($-r(t)-nc$), and subdiagonal elements, defined as
($r(t)$):

\begin{equation}\label{At}
	\hat{A}(t)=\left( \begin{matrix}
		-r(t)& 0   &  & \ldots  &  &   \\
		r(t)  & -r(t)-c   & 0 &  & &    \\
		0		& r(t)  & -r(t)-2c & 0 &   &   \\
				& 0 &   r(t)  & -r(t)-3c & \ddots & \\
		\vdots	&  	& 0 &  r(t)& \ddots & 	\\
		  &    &  &    & \ddots  
	\end{matrix} \right) \text{ .}\\
\end{equation}

The exact solution for $\vec{P}(t)$ is obtained following the procedure shown in Section~SI.3. However, this solution leads to a complicated and impractical survival probability after replacing $P_n(t)$ in Eq.~(\ref{Gsurv0}). Simple mathematical models are preferred in radiobiology for their ease of interpretation, computational efficiency, and ability to expedite treatment planning. They help biologists and clinicians understand treatment rationale, making them trusted tools. Thus, we aimed to simplify results using a first-order approximation in the Magnus expansion as

\begin{equation}\label{Geq}
        \vec{P}(t)=\exp\left( \int_0^t dt' \hat{A}(t') \right)  \vec{P}(t=0).
\end{equation}

\noindent Within this approximation, the transition matrix was integrated as $A_t=\int_0^t\hat{A}(t')dt'$, and $e^{A_t}$ was then computed. By multiplying this by the initial state $\vec{P}(0)=(1,0,0,\ldots,0)^T$, the resulting probability vector in Eq.~(\ref{Geq}) corresponds to the column vector of $e^{A_t}$, with each element given by:

\begin{equation}
	P_n(t)=\exp\left(-\int_0^t r(t')dt'\right) \left(\frac{\int_0^t r(t')dt'}{ct}\right)^n \frac{\left( 1-e^{-ct}\right)^n}{n!}, 
\end{equation}

\noindent (see Section~SI.3 for details). After replacing $P_n(t)$ in
Eq.~(\ref{Gsurv0}), the survival probability gets a functional form
that resembles the simplicity of Jung's model

\begin{equation}\label{Srepairt}
	\mathcal{S}(t)=\exp \left\lbrace  \frac{\int_0^t r(t')dt'}{ct}\left[
	1-ct-e^{-ct}\right] \right\rbrace.
\end{equation}

Since it is impossible to calculate all the contributions from an
infinite series, we assessed the error of the aforementioned
approximation as the contribution of the first order, assuming that
each next order contributes less and less for a repair function that
is saturating to a constant
value. %Moreover, if the parameter $k$ is small relative to treatment time, the time-dependent regression rate saturates fast to a constant value, and the error does not increase with treatment time.
Therefore, we calculated the error of $\vec{P}(t)$ as the difference
between the probabilities at first and second order. This results in
small error values when the parameters are almost constant throughout
treatment. As an example, using the Umodel's parameters obtained for
UT-SCC-14 cells under 44.5 $^\circ$C HT, the error values ranged below
$10^{-4}$ at $t = 1$ min and below $10^{-46}$ at $t = 60$ min (more
details shown in SI section SI.3, Eqs.
(SI.8)-(SI.15)). We also calculated the difference between the exact and approximated survival functions (Eqs. (SI.13) and (\ref{Sapp})) for this case, shown in Figure~SI.1. As illustrated in Figure~SI.1~b, the absolute error peaks at around 1.2\% near the 35-minute mark.\\

As radiation- and heat-induced damage is detected and restored by enzymatic mechanisms, we propose to model the regression process (in time) by means of an "effective'' enzymatic reaction representing the average kinetics of the restored molecules in the cell population. Detailed mathematical models have been proposed for some of the intracellular processes involved in DNA repair and protein refolding, introducing and numerically solving large sets of coupled differential/integral equations~\cite{Hall,Crooke,McMahon,Zheng,Ladjimi, Sivery,Scheff,Peper}. These frameworks lead to the desired regression rates but require numerous adjustable parameters, making those approaches unsuitable for practical survival models. Therefore, we introduced a source (damage) term into the Michaelis-Menten (MM) kinetics, an approach adapted to cover all cellular restoration processes based on common enzymatic reactions with a single average function. The underlying MM model consists of four differential equations describing the following enzymatic chemical reaction:

\begin{equation}\label{MME}
        [E]+[S]\rightleftharpoons_{k_r}^{k_f} [E \cdot S]\rightarrow_{k_{cat}}[E]+[P].
\end{equation}

Here, $[E]$, $[S]$, and $[P]$, denote the concentrations of the enzyme, the substrate (damaged molecule), and the product (restored molecule), respectively. $[E\dot S]$ refers to the concentration of enzyme-substrate intermediate complex. In this equation, the enzyme $E$ is associated to a substrate molecule $S$. This step occurs at rate $k_f$ but can also be reversed at rate $k_r$. After the interaction, the enzyme dissociates unchanged, and the substrate turns into a product molecule $P$. This part of the process occurs with a rate of catalysis $k_{cat}$~\cite{MMK}. After including the source of damage $r$ into the substrate and product equations, the time-evolution of the concentrations is given by: 

\begin{align}\label{MME2}
        \nonumber \frac{d[E]}{dt}&=-k_f[E][S]+(k_r+k_{cat})[E \cdot S]\\
        \frac{d[S]}{dt}&=-k_f[E][S]+k_r[E \cdot S]+r\\
        \nonumber \frac{d[E \cdot S]}{dt}&=k_f[E][S]-(k_r+k_{cat})[E \cdot S]\\
        \nonumber\frac{d[P]}{dt}&=k_{cat}[E \cdot S] -r[P].
\end{align}

Since the source term $r$ in the substrate kinetics is the rate of SLD production, it connects the enzymatic process with Jung's model. Under a quasi-steady state approximation, the enzyme-substrate complex is assumed to be constant $\frac{d[E \cdot S]}{dt}=0$. Accordingly, after upregulation, the total enzyme concentration remains constant ($\frac{d[E]}{dt}=0$) and equal to the initial value ($[E]+[E \cdot S]=[E]_0$). Under these conditions, the equation for the concentration of impaired molecules (substrate) reads

\begin{equation}\label{dSdt}
		\frac{d[S]}{dt}=-\frac{q_{max}[S]}{k'+[S]}+r.
\end{equation}

\noindent where $q_{max}=k_{cat}[E]_0$, and $k'=\frac{k_r+k_{cat}}{k_f}$. The solution of Eq.(\ref{dSdt}) for $r\le q_{max}$, assuming no impaired molecules $[S]_0=0$ at the beginning of the treatment is

\begin{equation}\label{qot2}
        [S]=\frac{k'}{q_{max}-r}\left[r+q_{max} W_0 \left( \frac{-re^{-y}}{q_{max}} \right) \right],
\end{equation}

\noindent where $W_0$ is the principal branch of the Lambert function, with $y=\frac{(q_{max}-r)^2t+k'r}{k'q_{max}}$. For the general case, we use Eq. (\ref{dSdt}) to approximate the concentration of impaired molecules up to first order in its Maclaurin series expansion as $[S]\sim [S]_0+[S]'_0t= [S]_0+rt$. This simplification allows for a more tractable model. The approximation holds under the condition of slow advance and regression rates ($tr,tq_{max} <k'$). \\
	
The net rate of molecular mending $\frac{d[P]}{dt}$ is obtained by subtracting the damage rate from the damage regression rate $\frac{d[P]}{dt}=q(t)-r[P]$. This means that the regression rate can be modeled as:
	
	\begin{equation}\label{qot}
		q(t)=k_{cat}[E \cdot S]=\frac{q_{max}[S]}{k'+[S]},
	\end{equation}
	
Under these assumptions, Eq.(\ref{qot}) becomes

\begin{equation}\label{rep}
        q(t)=\frac{d[P]}{dt}=\frac{q_{max}t}{k+t},
\end{equation}

\noindent with $k=k'/r$ being the average time to achieve the half of the maximum cellular capacity of mending impaired molecules, yielding a sigmoid rise of the repaired molecules over time. This is in line with the functional forms documented in~\cite{McMahon,Crooke}, but with the advantage of only requiring two adjustable parameters. The temperature dependence of the repair parameters adhering to the MM kinetics is explained as follows: the maximum/saturation value of the repair function $q_{max}=E_0 k_{cat}$, depends on the initial amount of repair enzymes $E_0$, which is influenced by the treatment stimuli, such as heat. The turnover number $k_{cat}$ reflecting enzyme efficiency and the time to achieve the half response $k=(k_r+k_{cat})/(k_f r)$ are also conditional on temperature, with each (average) chemical rate following an Arrhenius-type behavior. In this way, the parameters in the regression rate $q(t,T)$ can be linked to average temperature-dependent biochemical responses.\\

By inserting Eq.~(\ref{rep}) into Eq.~({\ref{rot}), the survival probability in Eq.~(\ref{Srepairt}) becomes:
	
\begin{equation}\label{Sapp}
\mathcal{S}(t)=\exp \left\lbrace  \frac{\left[ rt-q_{max}\left[t+k\ln\left(\frac{k}{k+t} \right)
	\right] \right] }{ct}\left[
    1-ct-e^{-ct}\right] \right\rbrace .
\end{equation}

\noindent Notably, the treatment time can be exchanged by the total dose ($t \rightarrow D$) in Eq.~(\ref{Srepairt}), given that the dose rate is constant throughout the treatment. This variable substitution only changes the units of the adjustable parameters $r,c,q_{max}$, and $k$.\\

At first glance, our model with a regression rate closely resembles the Multi-Hit-Repair (MHR) model of Scheidegger et al.~\cite{Scheidegger}, which describes the effects of RT and HT-induced radiosensitization (notably, the MHR model is not designed to reflect HT mono-treatment). However, the mathematical concepts defining doses and rates differ critically. In the MHR model, RT and HT doses are represented by state variables $\Gamma$ and $\Lambda$, proportional to dose rate ($\dot{D}=R$) and repair protein damage ($k_1$), counteracted by dose-dependent repair rates ($\gamma$ and $k_2$), respectively. The SLD accumulation rate is proportional to $R$, with a constant damage fixation rate. In contrast, our model specifies a constant SLD advance rate $r$ and a fixation rate $c$ proportional to the SLD accumulation stage $n$. Furthermore, in the MHR approach, the probability of repair from RT and TRT damage decreases exponentially with dose. On the contrary, in our model, repair is upregulated by treatment intensity, saturating at a maximum value. This unique feature of our model, among others, sets it apart from the MHR model, providing a comprehensive and distinct approach to describing the clonogenic survival curve flattening at higher doses.
\vspace{0.2cm}

\subsection*{Multiparametric optimizations}
\vspace{0.2cm}

We utilized the versatile non-linear least-square minimization of the python package lmfit~\cite{NewSteAllIng2014} to fit the necessary parameters of each model, i.e., the radiosensitivities ($\alpha$, $\beta$) of the LQ-model, the advance and damage fixation rates ($r$ and $c$) of Jung's model, and the parameters ($r$, $c$, $q_{max}$, and $k$) of the Umodel (see Eq.~(\ref{Srepairt})). All values were adjusted using the Levenberg-Marquardt algorithm~\cite{Levenberg,Marquardt} to fit the corresponding biological effect ($-\ln(S)$), experimentally obtained from the survival assays. The parameters are determined by minimizing the residuum $\epsilon$ of the biological effect 

\begin{equation}
        \epsilon=\sum_i\left[\ln{S_i}-\ln{f_S(x_i,\left \lbrace \gamma
            \right\rbrace )} \right]^2.
\end{equation}

Here, $i$ is a label for each survival probability $S_i$ in the experimental data set, $\left \lbrace \gamma \right\rbrace$ is the set of adjustable parameters, and $f_S(x_i,\left \lbrace \gamma \right \rbrace)$ is the corresponding prediction of the survival probability from the applied model. We set reasonable boundaries for the parameters, i.e., typically $r[\text{min}^{-1}]\in[0,50]$, $c[\text{min}^{-1}]\in[0,1]$, $Q_\text{max}[\text{min}^{-1}]\in(0,50]$, $k[\text{min}]\in(0,1]$, $\alpha[\text{Gy}^{-1}]\in[0,1]$ and $\beta[\text{Gy}^{-2}]\in[0,1]$. Note that, while any fit for $r$ and $c$ in Jung's model is in principle also valid for the Umodel with $q_\text{max}=0$, we explicitly restricted the minimization to different parameter values ($q_\text{max},k>t0$). The parameters' standard error (estimated 1$\sigma$ error bar) is also obtained from the optimization, and reported (Tables~SI.9, SI.10).\\

We must emphasize that the existence of several local minima for the error function hinders the search for a global solution. Thus, different values of adjustable parameters may produce similar values of $R^2 \sim 1$. This disadvantage is called ``lack of identifiability'', a significant problem in mathematical models of biological systems~\cite{Phan,Munoz,Alahmadi}. It is important to note that lack of identifiability is not the same as \emph{overfitting}. Overfitting occurs when a model is too complex and includes parameters that are unnecessary to represent the data accurately. For instance, the parameters $\alpha$ and $\beta$ of the LQ model are very identifiable, while Jung's model lacks identifiability despite the same number of adjustable parameters. In our model, the lack of identifiability is inherited from Jung's original model. To help overcome this problem, we restricted the solution space to parameters that satisfy the thermodynamic prediction described in~\cite{DeMendoza}. This condition improves the identifiability of the parameters in HT but does not fully solve it. As highlighted in the results section, the thermodynamic condition states that the SLD rate should grow exponentially with treatment temperature. Hence, for the Umodel in HT, $r$ is also restricted to depend exponentially on temperature. Moreover, we presumed that the maximum repair rate $q_{max}$ follows the trend of the inflicted damage. Accordingly, to reduce the ambiguity of the fitted parameters, we similarly imposed the exponential condition on $q_{max}$, for which the model reproduces the experimental data equally well. To account for these restrictions, the parameters at all temperatures $T$ are optimized simultaneously for each cell line, and the deviation of the functions $r(T)$, $q_{max}(T)$ from linear fits $\log[r(T)] = b (T-T_g)$, $\log[q_{max}(T)] = b^* (T-T^*_g)$, is added to the residuum. This way, the parameters $r$ and $q_{max}$ do not have to exhibit perfect exponential dependencies, which would be the case if they are directly replaced by exponential functions with the parameters $b$, $T_g$, $b^*$, and $T^*_g$ in the optimization. Instead, matching the fitted $r$, $q_{max}$ with exponential functions in terms of the corresponding coefficient of determination $R^2$ serves as an additional quality control of the model assumptions (see Table~SI.8 and Figures~SI.3, SI.8, and SI.9).\\ %(Table~\ref{exp}).\\

In all the cases, the resulting goodness-of-fit is reported by the coefficient of determination $R^2$ concerning the logarithm of the survival fraction. In addition, the corresponding Akaike information criterion $AIC$ is reported to account for the impact of the degrees of freedom. (see Tables~SI.9, SI.10). Note that due to the exponential dependencies of $r$ and $q_{max}$, the model has effectively fewer degrees of freedom when fitting survival fractions at several different temperatures. For instance, when fitting data at $N_T$ temperatures, there are effectively $4+2N_T$, instead of $4N_T$, fit parameters. Like other multiparametric optimization methods, the Levenberg-Marquardt algorithm is an iterative procedure that depends on the initial estimate of the parameter set $\left\lbrace \gamma\right\rbrace_0 $ and only converges to the global minimum if the initial estimate is already close to the solution.

\subsection*{Experimental methods}\label{crit}
\vspace{0.2cm}

\subsubsection*{Cell culturing}

Eight human HPV-negative HNSCC cell lines were applied in this study: SAS and HSC4 (HSRRB/JCRB, Osaka, Japan), UT-SCC-5, UT-SCC-14, and UT-SCC-60A (University of Turku, Finland), Cal33 (DSMZ, Germany), XF354 (DKFZ, Germany), and a subline of the FaDu-ATCC HTB-43 model (Dresden, Germany)~\cite{Eicheler}. Before use, the cell lines' genetic profile was verified via microsatellite analyses at the Institute of Legal Medicine (TU Dresden, Germany). They were also routinely tested free of mycoplasms using a PCR Mycoplasma Kit (AppliChem, Darmstadt, Germany), as detailed earlier~\cite{Chen}. The cell cultures were grown from validated frozen stocks for >2 to 620 passages (<120 cumulative population doublings) and cultured in standard Dulbecco's Modified Eagle Medium (DMEM) with L-glutamine, D-glucose (1 g/L) and 25 mM HEPES supplemented with 10\% heat-inactivated fetal calf serum (FCS) and 1\% penicillin/streptomycin (10,000 U/mL/ 10 mg/mL). Cells were kept in a humidified air atmosphere with 8\% CO$_2$ at 37 $^\circ$C. All culture media, supplements, solutions, and buffers were purchased from PAN- Biotech (Aidenbach, Germany). 

\subsubsection*{Colony formation assay (CFA)}

Exponentially growing cultures were enzymatically dissociated using 0.05\% trypsin/0.02\% EDTA in phosphate-buffered saline (PBS) to obtain single-cell suspensions. A CASY$^\text{\textregistered}$ TTC analyzer (Roche Innovatis, Reutlingen, Germany) was used to monitor cell culture quality and assess cell numbers and volumes in the single-cell suspensions for further use. Cells were then diluted appropriately and seeded in 6-well plates in cell line- and treatment-dependent concentrations using 1 ml of supplemented DMEM per well. Cells were incubated at 37 $^\circ$C for 20-24 h (less than one culture doubling for all of the cell lines) to allow adherence and overcome a potential proliferative lag phase due to the dissociation procedure. Plates were then exposed to HT and/or RT. After completion of treatment, 1 ml of supplemented DMEM medium was added to each well for extended culturing at standard conditions. The culture period for colon formation ranged between 7 and 14 days, according to $\geq 5$ cell line-specific culture doublings. The colonies were then washed with PBS, fixed for 10 min with 80\% ethanol followed by staining with a Coomassie blue solution. Colonies with $\geq 50$ cells were manually counted at low magnification to determine plating efficiencies and calculate survival fractions (S) of the treated samples relative to untreated controls. The choice of CFA setup is briefly discussed in Section~SI.4. All data used in the present study derive from $N\geq 3$ independent experiments with $n=3$ biological repeats.

\subsubsection*{Implementation of treatment: Hyperthermia and irradiation}

All hyperthermia treatments were performed using a pre-heated temperature-controlled PST-60HL-4 Plate Thermo-Shaker (BioSan, Latvia). The 6-well plates were transferred into the pre-heated shaker for defined times at temperatures of 40.5$^\circ$C to 46.5$^\circ$C, comprising the entire treatment period - from placing the plates in the device to removing them. As a prerequisite,  heating profiles in selected wells were recorded for different temperature settings via a TC-08 8-channel thermocouple data logger (Pico Technology, UK) combined with type T thermocouples (RS Components, UK)  before using the system for standardized HT treatment, according to~\cite{Chen3} to confirm that the target temperature in the 2-D culture setting is reached  within a few minutes and the cooling period is negligible. Control plates were incubated in parallel at 37$^\circ$C in the standard incubator. In the HT+RT treatment regimes, cells were irradiated at room temperature with 0 - 6 Gy single dose X-rays (200 kV; 0.5-mm Cu filter, approx. 1.32 Gy/min; YxlonY.TU 320 (Yxlon.international, Germany)) applied directly after completion of exposure to HT. The RT mono-treatment data used for modeling were acquired similarly but have been published previously~\cite{Chen}.

\subsubsection*{Inclusion criteria for experimental data sets from the literature}

Cell survival curves as a function of the applied dose were included only if they presented at least five experimental data points per curve. For RT, we exclusively extracted survival curves displaying dose-dependent flattening from the referenced literature. In these data sets, the dose needed to be given in Gy and the irradiation power reported in the original articles. For HT, the dose had to be expressed as treatment time for at least three different temperatures. Experimental points and error bars (when reported) from each data set were extracted using \emph{WebPlotDigitizer}$^\text{\textregistered}$. The names of the cell models follow the Cellosaurus nomenclature. Details of all data from the literature used in this study are given in Table~\ref{data}.

\section*{Results and Discussion}

In this work, we have developed a mathematical model to best describe clonogenic survival upon both radiation and hyperthermia mono-treatments and their combination. Our research shows that Jung's approach, initially designed to model cell survival after heat exposure, is also suitable for reproducing clonogenic survival curves after single-dose irradiation. This finding suggests that Jung's model could be a versatile tool for predicting both therapies' outcomes based on the accumulation of sublethal damage regardless of the energy type and a specific underlying cell death mechanism. However, Jung's basic model does not comprise a component of cellular recovery, which is essential to delineate the impact of heat on proteins of the DNA damage repair machinery. Therefore, as detailed in the Methods section, we propose a modified Jung's model (termed Umodel) that mathematically incorporates a regression rate. We applied the Umodel to clonogenic dose-response survival curves recorded in our laboratory for several HNSCC cell lines exposed to RT and HT treatments. Overall, the U-model demonstrates an improved performance, particularly in reflecting and extrapolating the survival of cell populations that display changes in cellular recovery with increasing dose manifested as straightening or flattening of the clonogenic cell survival curves. To emphasize this feature further, we fitted the Umodel to additional data reported in the literature (cf. Table~\ref{data}). The subsequent subsections present the results of multiparametric optimizations, demonstrating at least one set of parameters that lead to good performance in each case. The goodness-of-fit of the proposed Umodel is compared with the original Jung’s models as well as the LQ and the LQC model in the HT and RT mono-treatment scenarios (Sections~SI.5 and SI.6). For the latter two, the sign of the quadratic and cubic term is explicitely not restricted, such that these models are capable to reflect the straightening of survival curves~\cite{Bodgi}. More specifically, the LQC model often performs slightly better in data fitting but seems not plausible for predictions based on extrapolations. In the last part of the presented work, we applied the Umodel to experimental clonogenic survival data from two HNSCC cell lines exposed to HT immediately followed by RT, demonstrating the potential of the Umodel for modeling of combined TRT schemes (Section~SI.7). 
\vspace{0.2cm}

\subsection*{Rationale and performance of the Umodel in HT and RT mono-treatment modeling}
\vspace{0.2cm}

In experimental radiotherapy, the LQ function stands as a stalwart, providing a robust approach to modeling the survival fraction as a function of the irradiation dose~\cite{Brenner}. However, in the widely accepted mechanistic explanation by Chadwick and Leenhouts~\cite{Chadwick}, the LQ-model parameters ($\alpha$ and $\beta$) represent the appearance and accumulation of DNA double-strand breaks, which are not directly inflicted upon  HT exposure. Despite the LQ model's simplicity and  good performance, this discrepancy hinders its translation to the cellular survival processes under HT. Similarly, other mechanistical approaches, such as the Repair-Misrepair model~\cite{Tobias}, the Local Effect Model (LEM)~\cite{Friedrich4}, and the Giant Loop Binary Lesion model~\cite{Friedrich2}, are also not able to propose a biological principle that could be generalized to describe damage caused by both HT and RT mono-treatments. In contrast, considering an underlying mechanism of SLD accumulation, Jung's model promises a more general approach to describing the damage induced by any treatment. This untapped potential of Jung's model is an encouraging avenue for further exploration.\\

For the hyperthermia treatment, we experimentally determined the clonogenic survival curves in eight HNSCC cell lines as a function of HT exposure time at three different temperatures: 42.5~$^\circ$C, 44.5~$^\circ$C, and 46.5~$^\circ$C (Figure~\ref{RTHTown}a, Figure~SI.2). In principle, all four models - Jung's, LQ, LQC, and Umodel - perform similarly well, except for the UT-SCC-14 cell line at 44.5~$^\circ$C (see Table~SI.2, for $R^2$, \emph{AIC} and parameters' uncertainties). UT-SCC-14 cells present a significantly different behavior at this temperature than the other cell models, as the survival curve critically flattens with longer HT exposure times (Figure~\ref{RTHTown}a). In this case, the Umodel achieves better results. These observations indicate that the dose-dependent regression rate is not essential for curves with typical shoulders, and a constant SLD rate suffices. However, it is crucial for populations that show reduced cytotoxic effects as doses rise.\\

Since different sets of parameters in the Umodel lead to good fitting, we used a thermodynamic condition that restricts the rates $r$ and $q_{max}$ to grow exponentially with increasing temperature~\cite{DeMendoza}. As described in the Methods section, this condition aids the fitting and offers insight into some underlying phenomena and correlations without increasing the number of adjustable parameters of the model. For most HNSCC cell types, the difference between the damage and regression rates increases with temperature, and the curves start deviating as exemplified for the Cal-33 cell in Figure~\ref{RTHTown}b (other cell line data are shown in Figure~SI.3; the $q_{max}$ values are listed in Table~SI.2).\\

Figure~\ref{RTHTown}c presents the clonogenic survival with all fittings (LQ, LQC, Jung's and Umodel) for our previously published data sets obtained from the same eight HNSCC cell types treated with 200 kV X-rays instead of HT~\cite{Chen}. In these cases, the Umodel again gives coefficients of determination comparable to the other models (see detailed results, $R^2$, \emph{AIC} and parameters' uncertainties in Table~SI.3). Since repair speed and saturation might be encoded in the slope of the survival curves, we also tested the Umodel for the experimental data from Wells and Bedford~\cite{Wells}, who recorded RT survival curves for C3H/10T1/2 cells using three different radiation dose rates (0.49 Gy/h, 2.4 Gy/h, 55.8 Gy/h). The Umodel again demonstrated excellent fit results comparable to the benchmark models. To adjust the Umodel, we made assumptions based on the numbers of damaged target molecules and repair/response-associated mechanisms. Here, it is important to note that the common variances in the dose rates documented in preclinical therapy experiments and in clinical routine procedures when delivering individual fractions (1 - 5~Gy/min) do not affect the biological kinetics of DNA damage induction and repair~\cite{RBbook}. This differs from the application and study of real low-dose-rate irradiations~\cite{RBbook} and ultra-high-dose rate FLASH radiotherapy~\cite{Nikitaki2022}, which is beyond the scope of the present study but might be subject to future mathematical modelings.\\

\begin{landscape}	
	\begin{table}%[tbp]
		\centering
		\caption{\label{data} Data set details including cell type, treatment strategy and literature source}
%	\begin{adjustbox}{max width=\textwidth, max height=\textheight} 
\begin{adjustbox}{width=1.55\textwidth}
			\begin{tabular}{clccccc}
			\toprule
			\multicolumn{7}{c}{\textbf{Used data sets of RT or HT individually applied}} \\
			\midrule
			\midrule
			\multicolumn{1}{c}{\multirow{2}[2]{*}{Source}} & \multicolumn{1}{c}{\multirow{2}[2]{*}{Cell line}} & \multicolumn{1}{c}{\multirow{2}[2]{*}{Entity (cell type)}} & \multirow{2}[2]{*}{Treatment} & \multirow{2}[2]{*}{HT temperatures [$^\circ$C]} & \multicolumn{2}{c}{Displays flattening} \\
			&       &       &       &  & RT   & HT \\
			\midrule
			& Cal33 & human HNSCC & HT,RT & 42.5, 44.5, 46.5 & no    & no \\
			& HSC4  & human HNSCC & HT,RT & 42.5, 44.5, 46.5 & no    & no \\
			& SAS   & human HNSCC & HT,RT & 42.5, 44.5, 46.5 & no    & no \\
			& FaDu  & human HNSCC & HT,RT & 42.5, 44.5, 46.5 & no    & no \\
			\multicolumn{1}{p{8.355em}}{Own experiments} & XF354 & human HNSCC & HT,RT & 42.5, 44.5, 46.5 & no    & no \\
			\multicolumn{1}{p{8.355em}}{} & UT-SCC-5 & human HNSCC & HT,RT & 42.5, 44.5, 46.5 & no    & no \\
			%& UT-SCC-8 & human HNSCC & RT & - & no    & no \\
			& UT-SCC-14 & human HNSCC & HT,RT & 42.5, 44.5, 46.5 & no    & at 44.5 $^\circ$C \\
			%& UT-SCC-45 & human HNSCC & HT,RT & 42.5, 44.5, 46.5 & no    & no \\
			& UT-SCC-60A & human HNSCC & HT,RT & 42.5, 44.5, 46.5 & no    & no \\
			\midrule
			\cite{Jung} & CHO & chinese hamster ovary   & HT    & 40, 41, 41.5, 42, 42.5, 43.5, 44, 44.5 & -     & no \\
			\cite{Gerner} & Hela & human cervical carcinoma  & HT    & 41, 42, 43, 44, 45 & -     & at 41, 42, 43 $^\circ$C \\
			\cite{Ohara} & CFU-MG & murine bone marrow & HT & 41.8, 42, 42.3, 42.5, 43, 44 & -     & at 42, 42.3 $^\circ$C \\			 &  & granulocyte–macrophage progenitor &  &  &      &  \\
			\cite{Armour} & CCD-18Lu & normal human lung fibroblasts  & HT    &  41, 43, 45 & -     & at 41, 43, 45 $^\circ$C \\
			\cite{Armour} & WiDr & human colon carcinoma & HT    &  41, 43, 45 & -     & at 41, 43 $^\circ$C \\
			\cite{Armour} & A549 & human lung carcinoma & HT    &  41, 43, 45 & -     & at 41, 43, 45 $^\circ$C \\
			\cite{Armour} & U87MG & human glioblastoma-astrocytoma & HT    &  41, 43, 48 & -     & at 43 $^\circ$C \\
			\cite{Mackey} & CHO & Chinese hamster ovary & HT    & 41.5, 42, 42.5, 43, 43.5, 44, 44.5 & -     & at 41.5, 42, 42.5 $^\circ$C \\
			\cite{Sapareto} & CHO & Chinese hamster ovary & HT    & 42.2, 42.3, 42.4, 42.5 & -     & all \\
			\cite{Wells} & C3H/10T1/2 &  mouse embryonic & RT (Xray) & - & no & - \\
& & spontaneously immortalized cell line & & & & \\
			\cite{Habermehl} & HepG2 & human hepatoblastoma & RT ($^{12}$C) & -     & yes   & - \\
			\cite{Habermehl} & HepG2 & human hepatoblastoma  & RT ($^{16}$O) & -     & yes   & - \\
			\cite{Habermehl} & HUH7 & hepatocellular carcinoma & RT ($^{12}$C) & -     & yes   & - \\
			\cite{Habermehl} & PLC & hepatocellular carcinoma & RT ($^{12}$C) & -     & yes   & - \\
			\cite{Yagi} & SW1353 & human chondrosarcoma & RT ($^{12}$C) & -     & yes   & - \\
			\cite{Yagi} & HDF & normal human dermal fibroblasts & RT ($^{12}$C) & -     & yes   & - \\
			\cite{Weyrather} & CHO-xrs-5 & X-ray hypersensitive mutant of CHO & RT ($^{12}$C)    & -     & yes   & - \\
			\cite{Bruningk1} & HCT116 & human colon cancer & RT (Xray), HT    & 45, 46, 47  & no   & no \\
			\cite{Dikomey} & CHO & chinese hamster ovary & RT (Xray), HT & 42.5, 43  & no   & no \\
			\bottomrule
		\end{tabular}%
		\end{adjustbox}
	\end{table}
\end{landscape}

\begin{figure}%[H]
	\centering
	\includegraphics[width=1.0\textwidth,trim={0.9cm 20.5cm .9cm 0.5cm},clip]{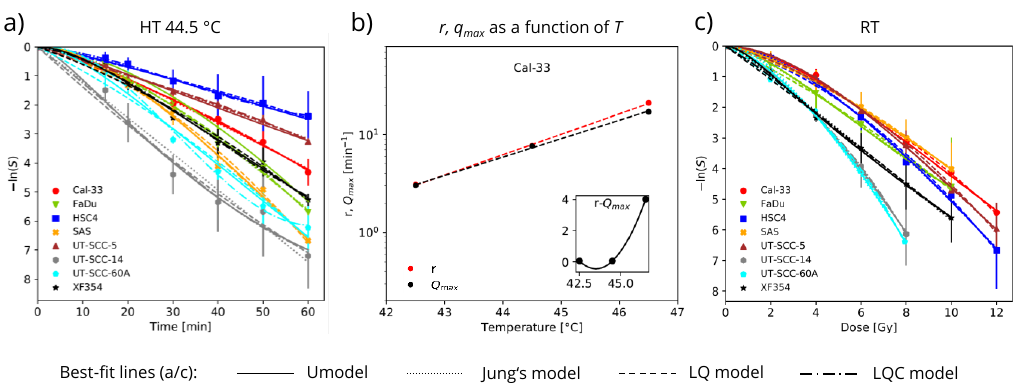}
	\caption{{\bf The Umodel shows comparable performance to the Jung’s, LQ, and LQC models to resemble clonogenic survival in HNSCC cells upon HT and RT mono-treatment.} Symbols represent cell survival fractions ($-\ln(S)$) of eight human HNSCC cell lines exposed to a) HT at 44.5$^\circ$C or c) 0-12 Gy single dose Xray~\cite{Chen}. The experimental, clonogenic survival curves for 42.5$^\circ$C and 46.5$^\circ$C are reported in Figure~SI.2. Data show means ($\pm$ SD) from $N = 3$ independent experiments with the models' best-fit lines. The coefficients of determination $R^2$, $AIC$ values and parameters' uncertainties are listed in Tables~SI.2 and SI.3. The advance rate in the SLD chain $r$ and the maximum repair rate $q_{max}$ determined from a) are     documented as function of temperature in supplementary Figure~SI.3 and exemplified for Cal-33 cells in b) with exponential fits (dashed lines); the inset displays the  difference between the two parameters on a linear scale.}
	\label{RTHTown}
\end{figure}

Based on these assumptions, we considered the dose rate as the sequential application of two small fractions within a finite interval, with $T$ as the target molecules (e.g., DNA) and $M$ as the repair/response molecules. The first fraction then damages the portion $T^*$ of $T$ and repairs $M^*$ of $M$. During the interval, the remaining ($M-M^*$) molecules initiate the damage response of $T^*$. With the second fraction, this process iterates. If no interval exists, both fractions concurrently damage $2A^*$ and $2M^*$ molecules, leaving fewer ($M-2M^*$) molecules to respond to more damaged targets 2$T^*$. This simplified approach suggests that sublethal damage ($r$) accumulates similarly in both treatment scenarios, and the maximum repair capacity ($q_{max}$) is equally affected. However, the activation time for the repair mechanisms ($k$) and the conversion rate from sublethal to lethal damage ($c$) increase due to frequent injury. Thus, we fixed the rates $r$ and $q_{max}$ in the Umodel and observed a monotonic increase in rates $c$ and $k$, achieving a goodness-of-fit coefficient of $R^2 \gtrsim 0.98$. Figure~SI.4 visualizes the results; all values, including $R^2$ and \emph{AIC} uncertainty parameters, are detailed in Table~SI.4.\\

Our model nicely reflects the flattening of the survival curve, as the sigmoidal regression function given by Eq.~(\ref{rep}) is upregulated and saturates to $q_{max}$ with increasing doses or treatment times. This behavior is not mimicked by models where the treatment doses reduce repair. The feature is particularly relevant for extrapolating and thus predicting treatment outcomes. Indeed, while all models quite well resemble any existing data points, some of them are expected to have poor predictive power. This limitation becomes evident in extrapolated fittings, such as those shown for HT in Figure~\ref{HTlit}a-c. Notably, the non-mechanistic LQC model, which in most cases seems to perform best in data fitting (see Tables~SI.2 to SI.3), is exceptionally poor in extrapolating beyond the existing data points (see for example curve fittings for UT-SCC-60 cells at 44.5 $^\circ$C and 46.5 $^\circ$C HT, SAS and FaDu cells at 46.5 $^\circ$C, or HSC4 and UT-SCC-5 cells at 42.5 $^\circ$C). In this context, we observed plausible results with the Umodel when applied to saturating (flattening) and non-saturating HT cell survival curves. In contrast, reasonably extrapolated fittings are achieved with all four models for the HNSCC clonogenic survival data upon RT; notably, none of the latter RT survival curves exhibit flattening (Figure~\ref{HTlit}d). Because of the Umodel's overall favorable performance, we emphasize its predictive potential for future validation and application.
\vspace{0.2cm}

\begin{figure}%[H]
        \centering
	\includegraphics[width=1.0\textwidth,trim={0.9cm 15.5cm .9cm 0.5cm},clip]{\sona 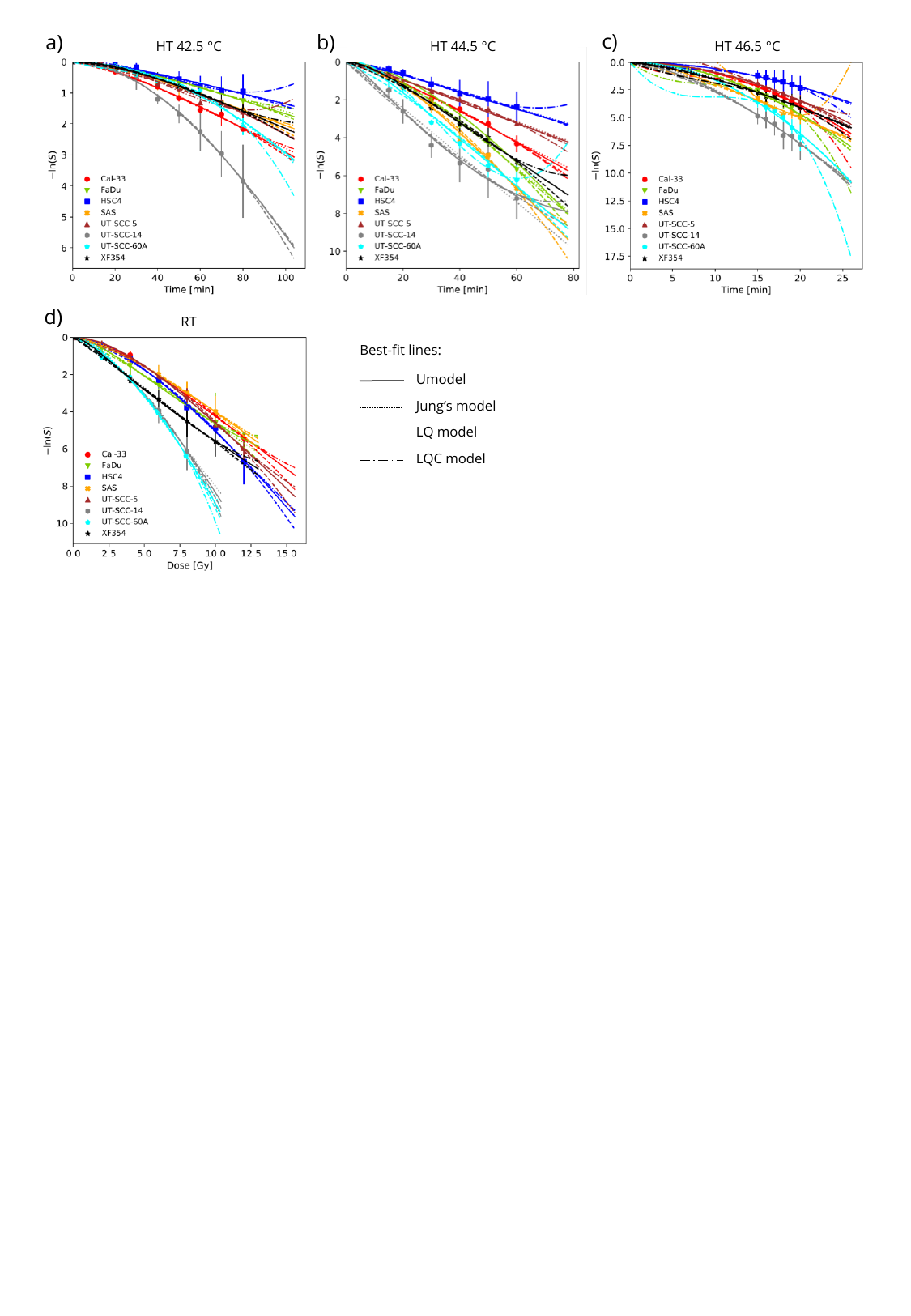}
        \caption{{\bf Extrapolations indicate plausible HT and RT
            survival outcome predictions with the Umodel but not the
            LQC model.} All model fittings from Figures
          \ref{RTHTown}a,c and SI.2 describing a-c) HT and d)
          RT responses in HNSCC cells are extrapolated up to 130\% of
          the total thermal/radiation dose and are represented by the
          extension of the best-fit lines beyond the last data point.}
        \label{HTlit}
\end{figure}

\subsection*{Emphasis on the peculiarity of survival curve flattening}
\vspace{0.2cm}

Our observation prompted us to test the Umodel further using more HT and RT survival data from the literature displaying the specific behavior of treatment response adaptation and clonogenic survival curve flattening. In principle, Jung's model fails to reproduce this type of data. The regression function introduced in the Umodel (Eq.~(\ref{rep})) critically improves the capability of the model to describe the referred behavior of the cellular population, such that the Umodel performs better than the LQ, the LQC, and Jung's model in those cases, see Table~\ref{adv}. Again, most prominently, the Umodel shows more realistic extrapolations beyond the reported experimental data. The extrapolated predictions are highlighted hereafter and their corresponding non-extrapolated fittings are documented in the supplementary material.\\

\begin{table}%[h]
        \centering
        \caption{\label{adv} Comparison of $R^2$ values for all cases displaying flattening in the clonogenic survival curves.}
\begin{adjustbox}{width=1.0\textwidth}
        \begin{tabular}{lclcccc}
                \cline{1-7}
                \multicolumn{1}{c}{Cell line} & \multicolumn{2}{c}{Treatment}
          &LQ model $R^2$&LQC model $R^2$& Jung's model $R^2$ & Umodel $R^2$ \\ \cline{1-7}
                CCD-18Lu ~\cite{Armour}   & HT& 41 $^\circ$C & 0.989 &
                                                                       0.992             & 0.846                             & {\bf 0.993}              \\
                A-549~\cite{Armour}  & HT& 41 $^\circ$C &0.932&{\bf 0.983}             & 0.824                             & 0.978              \\
                WiDr~\cite{Armour}  & HT& 41 $^\circ$C&0.878&{\bf 0.903}             & 0.827                             & 0.898              \\
                CCD-18Lu~\cite{Armour} & HT& 43 $^\circ$C&{\bf 0.998} &0.997
                                                              & 0.939
                                          & {\bf 0.998}              \\
                WiDr~\cite{Armour}  & HT& 43 $^\circ$C&0.986&0.988
                                                              & 0.872
                                          & {\bf 0.995}              \\
                A-549 ~\cite{Armour}     & HT&
                                               43 $^\circ$C&0.998&0.987             & 0.928                             & {\bf 1.000}              \\
                U87MG ~\cite{Armour}     & HT& 43 $^\circ$C&{\bf 0.999}&0.958             & 0.885                             & 0.997              \\
                A-549~\cite{Armour}     & HT& 45 $^\circ$C&0.995&{\bf 0.999}             & 0.984                             & {\bf 0.999}              \\
                CCD-18Lu~\cite{Armour}     & HT&
                                                 45 $^\circ$C&0.991&0.992             & 0.961                             & {\bf 0.995}              \\
                CFU-GM~\cite{Ohara}    & HT& 42 $^\circ$C& {\bf0.994} &{\bf 0.994}
                                                              & 0.984
                                          & {\bf 0.994}              \\
                CFU-GM~\cite{Ohara}      & HT&
                                                42.3 $^\circ$C&0.978&0.982           & 0.980                             & {\bf 0.987}              \\
                CHO~\cite{Jung}       & HT&
                                                41 $^\circ$C&0.991&{\bf 0.993}             & 0.992                             & 0.992              \\
                CHO~\cite{Mackey}     & HT&
                                               41.5 $^\circ$C&0.769&0.770           & 0.684                             & {\bf 0.801}              \\
                CHO~\cite{Mackey}     & HT& 42 $^\circ$C&0.961&0.968
                                                              & 0.659
                                          & {\bf 0.973}              \\
                CHO~\cite{Mackey}       & HT&
                                               42.5 $^\circ$C&0.956&0.967           & 0.884                             & {\bf 0.987}              \\
                CHO~\cite{Sapareto}   & HT&
                                                42.2 $^\circ$C&0.983&0.989           & 0.886                             & {\bf 0.995}              \\
                CHO~\cite{Sapareto}   & HT&
                                                42.3 $^\circ$C&0.985&0.985           & 0.861                             & {\bf 0.988}              \\
                CHO~\cite{Sapareto}   & HT&
                                               42.4 $^\circ$C&0.988&0.988           & 0.922                             & {\bf 0.991}              \\
                CHO~\cite{Sapareto}     & HT&
                                                 42.5 $^\circ$C&0.989&0.989           & 0.804                             & {\bf 0.991}              \\ \cline{1-7}
                Average                       & \multicolumn{2}{l}{} &0.966&0.969& 0.880                             & \textbf{0.976}              \\ \cline{1-7}
                & \multicolumn{1}{l}{} & \multicolumn{2}{l}{}              & \multicolumn{1}{l}{}        \\ \cline{1-7}
                \multicolumn{1}{c}{Cell line} &
                                                \multicolumn{2}{c}{Treatment}
          &LQ model $R^2$&LQC model $R^2$     & Jung's model $R^2$& Umodel $R^2$ \\ \cline{1-7}
                SW1353~\cite{Yagi}    & RT&  &0.999 &0.995
                                                              & 0.987
                                          & {\bf 1.000}              \\
                HDF~\cite{Yagi}   & RT&&0.996&0.990
                                         & 0.996
                                                              & {\bf 1.000}              \\
                XRS5~\cite{Weyrather}     & RT& &0.977&0.977
                                                              & 0.953
                                          & {\bf 0.982}              \\ \cline{1-7}
                Average  &                   \multicolumn{2}{l}{} &
                                                                   0.991
          & 0.987 & 0.979                             & {\bf 0.994}              \\ \cline{1-7}
        \end{tabular}
        \end{adjustbox}
\end{table}
%--------------------

\subsubsection*{Hyperthermia and thermal adaptation}
When cell cultures under HT become more resistant to increasing
thermal doses, the logarithmic clonogenic survival curves start
flattening~\cite{Dewey}. We call this behavior adaptation to
treatment. It is cell line-dependent and especially frequent for cells
exposed to mild HT. We experienced such behavior in some of our HNSCC
models, and similar observations come from several independent
literature data sets documenting clonogenic survival upon HT treatment
\cite{Jung,Gerner,Ohara,Armour,Mackey,Sapareto}. In these cases,
Jung's model reaches the limit of a straight line since it does not
include possible cell recovery and mitigation of thermal damage. This
is where our proposed modified model, the Umodel, comes in. It
reproduces clonogenic survival under these circumstances more
precisely, as visualized in Figure~\ref{fig:RTlit},
Figure~SI.10a, and Figures SI.5 to SI.7; Tables~SI.5 to SI.6 document the respective fit data. Note that in several cases the Umodel does not just achieve an higher $R^2$ but
also a lower $AIC$ value than the other three models, see Tables~SI.9~d,e indicating that the superior agreement with the experimental survival curves from hyperthermia treatment is not merely due to having 1-2 more model parameters. The corresponding increase in the rates $r$ and $q_{max}$ with temperature is reported in Figures SI.8 to SI.10.\\

We hypothesize that cell cultures exposed to moderate HT, particularly around $\simeq43^\circ$C, adapt to stressful conditions by upregulating survival mechanisms and enhancing recovery. Consequently, the decline in the survival rate slows down as the exposure time increases, and the expected shoulder of the survival curve ($-\ln(S)$) no longer takes place. Several studies have shown that the response to HT is triggered by protein denaturation, where heat shock proteins (HSPs) are activated, and heat shock factors (HSFs) are upregulated in a nonlinear manner~\cite{Zheng,Ladjimi,Sivery,Scheff,Peper}. Our work comprises, and mathematically simplifies, those regulatory mechanisms through a modified Michaelis-Menten model, capable of describing the nonlinear rise of the refolded proteins during the exposure time, as dictated by Eq.~(\ref{qot}). Introducing a mending rate (Eq.~(\ref{rep})) into Jung's model (Eq.~(\ref{Srepairt})) thus mathematically defines the adaptation to treatment.\\

\begin{figure}%[H]
        \centering
	\includegraphics[width=.95\textwidth,trim={1.9cm 21.5cm 1.9cm 0.7cm},clip]{\sona 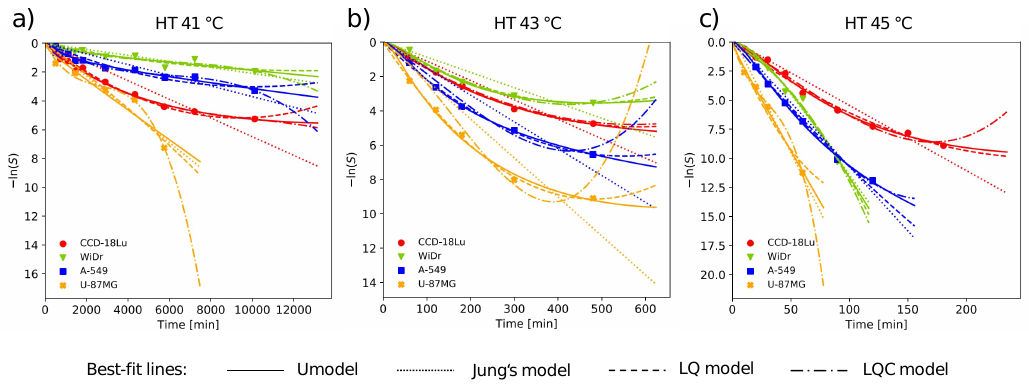}
        \caption{{\bf Extrapolations indicate plausible HT survival
            outcome predictions with the Umodel based on fittings to
            data extracted from literature.} Symbols represent cell
          survival fractions ($-\ln(S)$) obtained from clonogenic
          assay data extracted from reference \cite{Armour}; selected data
          (means $\pm$ SD) from different human cell types - normal
          human fibroblasts CCD-18Lu, lung carcinoma A-549,
          glioblastoma-astrocytoma U-87MG, and colon carcinoma WiDr
          exposed to HT at a) 41$^\circ$C, b) 43$^\circ$C, and c)
          45$^\circ$C~\cite{Armour} are shown. The data are fitted
          with the Umodel, Jung's, and the LQ and LQC models and
          extrapolated up to 130\% of the maximal treatment time. The
          extrapolated regions are represented by the extension of the
          best-fit lines beyond the last data point. The
          non-extrapolated curves are shown in
          Figure~SI.10a. The coefficients of
          determination $R^2$, $AIC$ values and parameters'
          uncertainties are listed in Tables~SI.6 and
          SI.9. Additional supporting data are documented in
          Figure~SI.9.}
        \label{fig:RTlit}
\end{figure}

\subsubsection*{Irradiation and the high-dose radioresistance phenomenon}
A meta-study from 2021~\cite{Friedrich} compared the outcomes of ion
beam irradiation with reference photon irradiation (X-ray), surveying
\textit{in vitro} clonogenic cell survival data across the literature.
The authors identified several experimental series showing signs of
cell resistance with higher radiation doses, i.e., the
linear-quadratic behavior at lower doses transitioning into a purely
logarithmic or flattening (saturating) dose-response relationship at
higher doses. The flattening in the survival curve is expressed as
negative $\beta$-values in the LQ model fittings. This may lead to
even concave line of best fit and also contradicts the mechanistic
interpretation of the radiobiological parameters in the LQ model
\cite{Chadwick}. The LQC model addresses this discrepancy by employing
an additional cubic term in the exponent of the LQ model. We,
therefore, next tested the LQ and the LQC models versus the Umodel in
three of such data sets extracted from published literature, ensuring
a thorough and meticulous process. Again, we discover the Umodel's
superior performance, particularly visible in the extrapolations,
indicative of putatively higher predictive power. The results,
documented in Figure~\ref{Combi} and Figure~SI.10b are a testament to the robustness of our approach. Table~SI.7 provide a comprehensive summary of all
parameters and $R^2$ values. In several cases the Umodel achieves also a lower $AIC$ value than the other three models, (see Table~SI.9b) indicating again that the superior agreement with the experimental survival curves from radiotherapy is not merely due to the increased number of parameter.\\

\begin{figure}[h!]
        \centering
	\includegraphics[width=.95\textwidth,trim={1.9cm 21.5cm 1.9cm 0.7cm},clip]{\sona 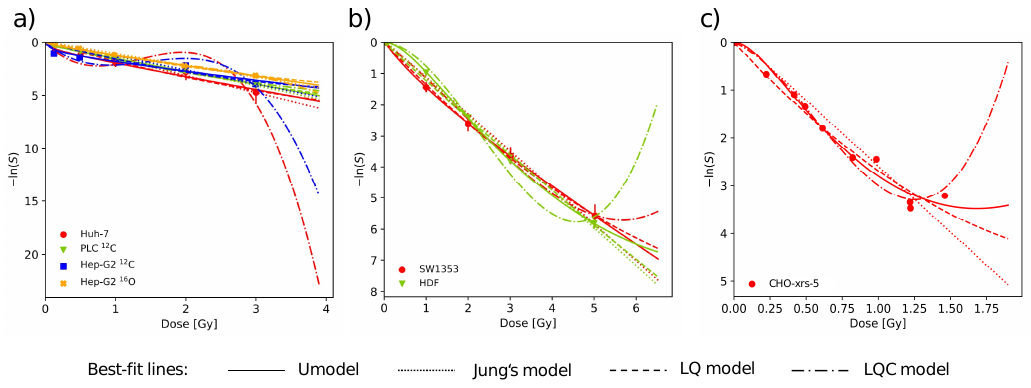}
        \caption{{\bf The Umodel shows better performance than the
            Jung’s, LQ, and LQC models in extrapolating flattening
            clonogenic survival curves upon particle RT.} Symbols
          represent cell survival fractions ($-\ln(S)$, mean $\pm$ SD)
          obtained from clonogenic assay data extracted from references \cite{Habermehl,Yagi,Weyrather} ; a) human hepatocellular carcinomas Huh-7 and PLC and hepatoblastoma HepG2 exposed to carbon ions $^{12}$C and hepatoblastoma HepG2 exposed to oxygen ions           $^{16}$O~\cite{Habermehl}. b) chondrosarcoma SW1353 and
          normal human dermal fibroblasts HDF exposed to carbon ions
          $^{12}$C~\cite{Yagi}, c) radiosensitive mutant of Chinese
          hamster ovary cells CHO-xrs-5 exposed to carbon ions
          $^{12}$C~\cite{Weyrather}. The data are fitted with the
          Umodel, Jung’s, and the LQ and LQC models and extrapolated
          up to 130\% of the maximal treatment time The extrapolated
          regions are represented by the extension of the best-fit
          lines beyond the last data point. The non-extrapolated
          curves are shown in Figure~SI.10b. The coefficients of determination $R^2$, $AIC$ values and parameters' uncertainties are listed in Tables~SI.7
          and SI.9.}
        \label{Combi}
\end{figure}

An early mechanistic interpretation of the high-dose radioresistance phenomenon in single-dose irradiation experiments suggests that cell subpopulations with different sensitivities co-exist. Here, the resistant subpopulations dominate clonogenic survival at higher doses, manifesting a "resistant tail'' of the survival curve~\cite{Denekamp}. The regression rate of the Umodel reflects such
scenarios to some extent by encompassing an average upregulation in the DNA repair capacity. A more recent alternative hypothesis by Friedrich et al.~\cite{Friedrich4} proposes a model based on the spatial distribution of the DSBs within a discrete organized chromatin region on a megabase pair scale - a giant loop. In this case, the deviation from the survival curve at higher radiation doses predicted via the LQ model is attributed to the formation of clustered DNA damage, defined as the mutual effect of DSBs over more considerable genomic distances. The model assumes the highest radiation efficiency if precisely two DSBs are induced within one loop. More than two DSBs on average per loop do not linearly enhance the radiation response. As a consequence, the relative contribution to lethality per DSB decreases, and a saturation effect occurs~\cite{Friedrich2}. This phenomenon may explain the lower effectiveness of higher doses as reflected by straight or flattening tails of the survival curves. Such a mechanism is expected to be more critical for high LET/particle irradiation, i.e., the probability of inducing cluster DNA damage is higher than for conventional X-rays~\cite{Friedrich2}. Variation in the LQ model $\beta$ values with LET have already been demonstrated~\cite{Friedrich,Friedrich3}. This particular mechanistic link has not yet been considered in our model. However, the reduced lethality at higher doses can still be modeled as a decreased advance rate $r$, equivalent to the more effective repair in the Umodel.\\

While not our primary focus, we shall emphasize that the radiation response curve flattening phenomenon is also observed in lymphocytes. Indeed, clonogenic survival assessment in peripheral blood lymphocytes has been a gold standard \textit{in vitro} assay for determining individual radiosensitivities in humans. However, response rates may also derive from distinct analytical endpoints and may not unequivocally overlap with clonogenic survival outcomes. A 2023 comprehensive review of in vitro and \textit{in vivo} studies in this field revealed that lymphocyte response curves based on different analytical endpoints show more shallow slopes and saturation at higher doses in most cases~\cite{Pag2023}. Pham \textit{et al.} recently presented a mathematical saturation model assuming a Poisson distribution of cell survival over DNA damage to better reflect the radiation response of lymphocytes~\cite{PhaCouThaVal2024}. This model performed better than the LQ model when surviving fractions were estimated from apoptosis (rate) detection. While this seems reasonable in lymphocyte response assessment, it is known that the survival of cancer cells upon radiation is not related to apoptosis induction. Therefore, our model development focuses on clonogenic survival curve flattening only, thereby avoiding the exclusion of any specific mechanism leading to permanent loss of reproductive capacity referring to radiotherapeutic “cell kill” as highlighted earlier (see Methods: Biological Terminology). 
\vspace{-0.2cm}

\subsubsection*{Survival curve flattening in the context of RT and metabolic targeting} 
The Umodel allows for mathematical simulation of different damage sources. Hence, it can also be considered for modeling the outcomes of simultaneous combination therapies. For demonstration, we next applied the Umodel to selected clonogenic cell survival data of a previously published study from our laboratory, where we identified cases of survival curve flattening when combining RT with a metabolic targeting strategy~\cite{Chen}. Here, a panel of HNSCC cell lines was deprived during clonogenic survival assessment of the proteinogenic amino acid arginine for 24 hours before and throughout irradiation and compared to RT alone. Jung's model and the Umodel fit most of the selected data series very well. However, those survival curves that display a flattening course at higher radiation doses are again reproduced more precisely by the Umodel and the LQC approach as opposed to Jung's initial and the LQ model (Figure~\ref{fig:extrapolation}, Figure~SI.11, Table~SI.10). Mechanistically, the decrease in the slope of the treatment outcome ($-\ln(S)$) for this simultaneous combinatorial treatment could be explained by the potential existence of subpopulations with different sensitivities and responsiveness to proteogenic and ER stress resulting from the extended lack of arginine. According to the RT and HT mono-treatments and based on the extrapolated clonogenic survival fittings, the Umodel appears better suited for predicting the response to combinatorial therapy beyond the existing data points than the LQC model.\\ 
\vspace{-0.2cm}

\begin{figure}[h!]%[!t]
	\includegraphics[width=.95\textwidth,trim={0.9cm 21.5cm .9cm 0.7cm},clip]{\sona 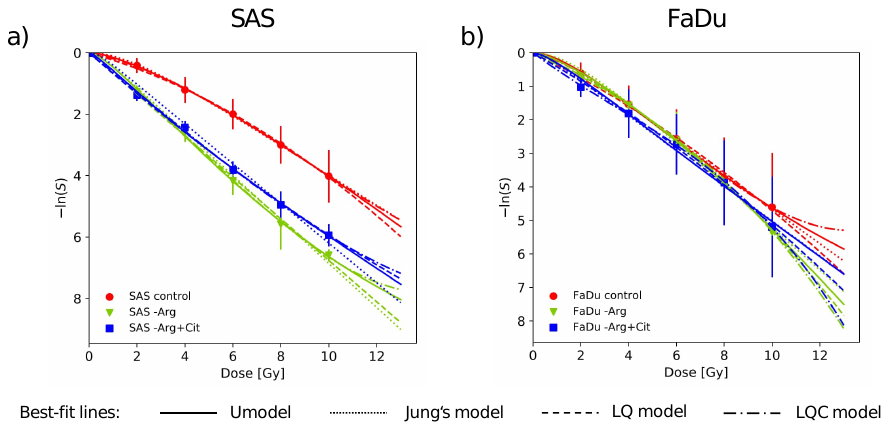}
	\caption{{\bf The Umodel performs well in extrapolating
            clonogenic survival upon simultaneous combinatorial RT
            employing different damage sources: an example of
            metabolic targeting therapy simultaneously applied with RT
            in two HNSCC cell lines.} Symbols represent cell survival
          fractions ($-\ln(S)$) obtained from clonogenic assays using
          a) SAS and b) FaDu HNSCC cell lines exposed to metabolic
          stress conditions, i.e., arginine deprived without (-Arg) or
          with citrulline enrichment (-Arg+Cit), combined with 0-10 Gy
          of single dose X-ray irradiation~\cite{Chen}. The data, extracted from reference ~\cite{Chen}, show means ($\pm$ SD) from $N = 3$ independent experiments and the best-fit lines of all mathematical models of interest. The fittings are also extrapolated up to 130\% of the maximal radiation dose. The extrapolated regions are represented by the extension of the best-fit lines beyond the last data point. The non-extrapolated curves are shown in Figure~SI.11. The coefficients of determination $R^2$, $AIC$ values and parameters' uncertainties are listed in Table~SI.10.}
	\label{fig:extrapolation}
\end{figure}

Taken together, survival curves that do not conform to the standard LQ model can appear upon HT and irradiation mono-treatments and simultaneous combinatorial RT. The pressing need for more generalizable mathematical models is underscored by the fact that the underlying phenomena also have profound relevance for treatment planning and prognosis in the clinical setting. The limitations of the current models in capturing the full complexity of these survival curves further emphasize the necessity of our proposed Umodel, which includes a regression function between consecutive stages of SLD accumulation, making it versatile and applicable for atypical clonogenic survival curves. In summary, our observations further demonstrate the versatility of the Umodel for generalized predictions that do not rely on a single or selective underlying biological mechanism. We, therefore, expect for our approach also to model the outcomes of simultaneous TRT. However, TRT is most frequently provided consecutively without or with a treatment gap between the individual modalities. As a consequence, more complex interrelations must be considered, requiring advanced mathematical combination treatment modeling, as will be briefly highlighted in the next chapter.\\

\subsection*{Considerations, challenges, and perspectives for applying the Umodel}
\vspace{0.2cm}

\subsubsection*{Modeling of combined thermoradiotherapy} 
The Umodel describes
the cell survival as the net rate of sublethal damage progression
(Eq.~(\ref{rot})) as a result of the dynamics between damage (increase
in $r$) and recovery (reduction in $q(t)$; lower $q_{max}$ or higher
$k$), regardless of a specific underlying mechanism. Therefore, in
principle the impact of any radiosensitizing agent that directly
affects the tumor cells could be modeled with our approach. In the
presented work, we demonstrate this point by modeling
radiosensitization induced by metabolic deprivation (Figure~\ref{fig:extrapolation}) and by applying heat, as briefly outlined below.\\

As a starting point for the sustained added value of our unique approach, we here demonstrate the first application of the Umodel to survival data of two HNSCC cell lines treated consecutively with HT and RT. Notably, these data sets do not show any flattening. Anyways, our mathematical approach, describing therapy outcome via SLD accumulation, has been adapted to encompass various biological aspects affecting the state of the cells and cell populations within the compartments for the TRT setting. The radiosensitizing efficacy of HT has been widely proven in different \textit{in vitro} and \textit{in vivo} models of various normal and cancer cell types~\cite{Overgaard,Vujaskovic,Konings,Oei2,Bruningk3,Chen3}. One of the most plausible mechanisms proven to explain, at least partially, the thermal enhancement of ionizing radiation is the impairment of the DNA-repair machinery that fix the radiation-induced damage. This additional synergistic effect might be the result of thermal denaturation of DNA-repair enzymes, particularly affecting the base excision repair (BER) and homologous recombination (HR) pathways, as observed in various mouse and human cell types~\cite{Oei,Mei,Kampinga1,Kampinga2}.\\

In the adapted Umodel, the parameters of the mono-treatments, i.e., RT and HT at different temperatures, are calibrated independently as before. Since the combined treatment in our data sets starts with hyperthermia, outcome is computed by integrating the set of ordinary differential equations of the Umodel Eq.~(\ref{BE}) with the parameters of hyperthermia mono-treatment. The proposed HT-RT synergistic effect mainly depends on the impact of HT on repair and is functionally implemented by the enhancement of sublethal damage of RT. From thermodynamic principles, linear and exponential relations of SLD augmentation with HT treatment time and temperature, respectively, are considered. Mathematically this is achieved by adjusting, i.e., reducing, the maximal RT repair rate $q_\text{max}$, affecting the probabilities of cells within successive compartments and the overall progress towards the death compartment, see Section~SI.7. More complex treatment schedules would require more advanced calculations. However, in principle, the Umodel will allow to model the outcomes for different treatment orders and is also considered to mimic the impact of treatment gaps which shall be the focus of future work.\\

According to the highlighted scheme, the Umodel is fitted to experimental clonogenic survival data of FaDu and SAS HNSCC cells which were first exposed to three different HT treatments (40.5 $^\circ$C, 42.5 $^\circ$C, and 44.5 $^\circ$C for 15-30 minutes) and then immediately thereafter irradiated with $0-6$ Gy single dose X-ray. Notably, the gap between applications in the combined treatment has effectively been null in all of these experiments. Averaged clonogenic survival curves from $N=3$ independent experiments and the respective Umodel fittings are documented in Figure~SI.12, showing the expected excellent performance; the fitting parameters are summarized in Table~SI.11. With rising temperatures, the model reflects the increase in the SLD rate as expected, but counterintuitively shows a decrease in the damage fixation rate of radiation (see Figure~SI.13e-f, and $c$ values in Table~SI.11). This observation is difficult to interpret but clearly leads to new mechanistic hypotheses to be addressed in future studies by specifically designed experiments.\\

\subsubsection*{A critical look at biological mechanisms and clinical translation} 

The parameters $r$ and $c$ in the Umodel are surrogates for the sum of biological processes involved in the response to hyperthermia (HT) and radiotherapy (RT), such as DNA breaks, protein denaturation, endoplasmic reticulum (ER) stress, heat shock response (HSR), and DNA repair. Specifically, r denotes the rate at which these processes cause potentially reversible non-lethal cellular damage. In contrast, $c$ represents the rate at which the same processes induce irreversible cell fazes, ultimately accompanied by proliferative cell death. Together, $r$ and $c$ offer a comprehensive view of how cellular damage transitions from repairable to lethal under HT and RT conditions take place. The parameters $q_{max}$ and $k$ are primarily biochemical indicators associated with the enzymatic recovery of cellular components. $q_{max}$ reflects the enzymatic capacity, while 1/$k$ indicates the rate at which these enzymes counteract cellular damage. These parameters encompass the initial capacity for pro-survival mechanisms ($q_{max}$), and the speed at which these mechanisms are activated ($k$). The parameters can be directly related to biological observations of cellular response to treatment. For example, we found cells in FaDu spheroids to more effectively activate pro-survival mechanisms upon combined HT+RT than SAS spheroids~\cite{Chen3}. FaDu cells have a higher basal expression of certain protective proteins (affecting $q_{max}$) and are able to more efficiently induce the expression of these proteins upon treatment (affecting 1/$k$). Notably, inherent and induced deficiencies in DNA repair are accounted for by the model through the parameters of the function $q(t)$ that describe the maximum capacity of enzymatic proteins involved in DNA and cell repair ($q_{max}$) and their average response time ($k$). Thus, distinct types of impaired responses are reflected by lower maximum capacities and longer response times relative to the rate of damage induction. Still, it is important to emphasize, that the interconnected nature of $r$ and $c$, as well as $q_{max}$ and $k$, creates an identifiability issue among these parameters, which limits the precision of their individual interpretation.\\ 

The main limitation of the proposed Umodel is the lack of identifiability of its parameters inherited from Jung's model. Lack of identifiability is quite a common problem in mathematical models and is receiving increasing attention in the applied mathematics community~\cite{Phan,Munoz,Alahmadi}. It follows from the fact that some parameters correlate with each other, and therefore, different sets of parameter values lead to very similar goodness-of-fit within the uncertainty of the experimental data. This flaw in the model hinders the interpretation of the specific parameter values and their association with the sensitivity of different cell types to treatment. Consequently, in this paper, we can describe trends of the parameters as a function of thermal or radiation dose but not yet the biological meaning of the specific parameter values. Nevertheless, the parameter trends can provide more insights into the underlying phenomena because they come from simultaneous fitting to several independent data sets. One way to improve the identifiability is to restrict the multiparametric space, as we have done by introducing thermodynamic conditions for the dependence of the model parameters on HT exposure time and temperature. However, the problem still needs to be fully solved; more specific experimental data describing DNA damage and repair would be helpful to validate our current assumptions and define a reasonable physiological range for the parameters, improving their identifiability. For instance, extensive experimental quantification of enzyme activities relevant to DNA repair or protein refolding upon treatment would help to refine the proposed model regarding its biological relevance and robust interpretability.\\

Like many others, we have
begun developing our model based on 2D clonogenic survival curves
obtained after single-dose irradiation. While the underlying assay can
reflect the intrinsic radiosensitivity of cells, it does not account
for factors related to the tumor microenvironment nor facilitate the
assessment of clinically relevant fractionated irradiation regimens.
We are aware of these limitations and plan to extend our mathematical
approach to include 3D and in \textit{in vivo} assays and clinically applied
treatment schedules. Ultimately, a mathematical model such as our
Umodel that more accurately represents the intrinsic treatment outcome
of cells - incorporating phenomena such as thermal adaptation and
radioresistance at high doses - will provide a significant advantage over currently used mathematical functions. The future goal is thus to advance this model for translation to \textit{in vivo} studies and eventually into clinical practice. In the clinical context, our approach allows using a single unified model based on generalized principles of accumulation of sublethal damage with implemented radiosensitization. Regardless of the type of energy deposited and mechanism of action, the Umodel can reproduce various patterns of clonogenic survival curves, including any flattening, thus encompassing the variability of cell reactions to therapy and thereby potentially better reflecting overall tumor responses. The features of the Umodel may also resemble radiation dose-dependent modifications in chemo-radiotherapy outcomes (see for example clonogenic survival curve patterns in~\cite{Chaudhary,Agoni}). This shall be proven in future applications. \\

\section*{Conclusions and Outlook}

In this work, we extended Jung's approach, which initially described clonogenic cell survival after HT, to also model radiation treatment outcomes and incorporate adaptation to therapy. Due to its compartmental structure, the developed unified model (Umodel) allows the accumulation of SLD without assuming or excluding any particular mechanism of injury. This feature and its mathematical formulation make the model suitable for describing/predicting the therapeutic outcome of the individual treatments (RT, HT) and their synergistic combination based on the same general principles. The thermodynamic condition for the HT dependence of the Umodel's parameters helps the fitting, supporting protein denaturation as a plausible explanation for radiosensitization. Our consistent approach opens a range of options for further model developments and strategic therapy outcome predictions, e.g., to account for differences in sequential treatments with intervals between them, which could not yet be implemented.\\

Since the Umodel model rests on the accumulation of non-lethal damage, it also naturally allows the inclusion of pro-survival mechanisms, modeled as effective enzymatic reactions. This characteristic is highly relevant to describe tumor cells that can adapt to treatment, e.g., some cell subpopulations under selected radiation treatments or mild hyperthermia. Our model of enzymatic restoration of damaged molecules contains several simplifications encompassing different processes such as DNA repair, protein refolding, or the redistribution of subpopulations due to heterogeneity and plasticity. Through this effective overall enzymatic reaction, the Umodel is able to reproduce and predict even atypical average outcomes of the entire cell populations by adding only two adjustable parameters $(q_{max}, k)$ to the Jung model. The high performance of the Umodel is stressed especially in cases presenting adaptation to treatment and flattening of survival curves; its predictive potential is indicated by fit extrapolations, where other models may fail. These assumptions shall be further validated in retrospective and prospective settings.\\

The Umodel describes effects as damage accumulation and death in cell cultures as a function of the heat or radiation dose, assessed by treatment time at fixed dose rates. Future extensions of the model could incorporate chronological time. For calibration, this requires additional analytical endpoints of cell damage and death after treatment that can be monitored over time, such as DNA damage and repair or factors reflecting regulated and non-regulated cell death processes. Moreover, our unique model possesses two crucial attributes that broaden its potential applications, which we plan to address in future work. Firstly, treatment is represented mathematically as an operator that modifies the initial state of the cell population. Secondly, it provides a closed expression for cell survival, which simplifies the derivation of biologically effective and equivalent doses, enabling exploration of fractionated treatment regiments. We envision that these features will facilitate outcome modeling and putative prediction of sequential treatments applied in different orders and varying recovery intervals between them. We further propose its direct incorporation into more complex mathematical models of multicellular dynamics, such as tumor spheroids, to represent \textit{in vivo}-like, more clinically translational outcomes~\cite{Franke}.
\vspace*{0.5cm}

\textbf{Authors contributions}

A.M.D.M. conceived the presented idea and developed the theory. S.M. and L.A.K-S designed and supervised the experiments. S.M. and L.E. performed the experiments. A.M.D.M., P.S.C, S.L., and A.G.M. performed the numerical calculations and analyses. All authors contributed to the interpretation of the results. A.M.D.M., S.M., S.L, and L.A.K-S designed and wrote the manuscript. L.A.K-S. supervised the project
\vspace*{0.5cm}

\textbf{Acknowledgments}

This work was supported by the German Federal Ministry of Education and Research (BMBF; 03Z1N512; 16dkwn001). We thank Ms. Marit Wondrak for technical support of the biological experiments and Dr. Damian D. McLeod and Dr. Oleg Chen for helpful discussion. We gratefully acknowledge the valuable contributions of the reviewers invited by Physics in Medicine and Biology.

%\printbibliography
\bibliographystyle{agsm} 
\def\newblock{\ }%
\def\BIBand{and}%

\end{document}